\documentclass{ws-rv961x669}
\usepackage{ws-rv-van}     % numbered citation/references (default)
\usepackage{ws-rv-thm}     % comment this line when `amsthm / theorem / ntheorem` package is used
\usepackage{subfigure}     % required only when side-by-side / subfigures are used
\makeindex
\begin{document}

\chapter{Neutrino Mass Ordering in Future \\ Neutrinoless Double Beta Decay Experiments} \label{ra_ch1}

\author[Jue Zhang]{Jue Zhang}

\address{Institute of High Energy Physics, Chinese Academy of
Sciences, \\ Beijing 100049, China\\E-mail: zhangjue@ihep.ac.cn}

\begin{abstract}
Motivated by recent intensive experimental efforts on searching for neutrinoless double beta decays, we present a detailed quantitative analysis on the prospect of resolving neutrino mass ordering in the next generation $^{76}$Ge-type experiments.

\end{abstract}

%\markright{Customized Running Head for Odd Page} % default is Chapter Title.

\body

%\tableofcontents

\section{Introduction} \label{sec:intro}

The quest for neutrinoless double beta decay ($0\nu\beta\beta$) has experienced an interesting history~\cite{Barabash:2011mf}. In 1939, combining the idea of double beta decay ($2\nu\beta\beta$) proposed by Goeppert-Mayer four years ago and the notion of Majorana fermions formulated by Majorana two years ago, Furry for the first time discussed the possibility of observing two electrons but without any neutrinos during the transformation of the $(A, Z)$ nucleus to the $(A, Z+2)$ nucleus, i.e., $(A, Z) \rightarrow (A, Z+2) + 2e^-_{}$. At that time, simply based on the phase space argument, this $0\nu\beta\beta$ mode was believed to have a half-life around $10^{15}_{}~{\rm yr}$, much shorter than that of the corresponding $2\nu\beta\beta$ mode, $\sim 10^{21}_{}~{\rm yr}$. If such an estimation were true, one would have discovered $0\nu\beta\beta$ quite shortly, maybe by the 1950s. 

The first turning point in the $0\nu\beta\beta$ history is the discovery of parity violation in 1957 and the establishment of V-A theory, which concludes that if neutrinos were massless, $0\nu\beta\beta$ would be exactly forbidden. Such a discouraging result declared that the experimental search for $0\nu\beta\beta$ entered the ``middle ages"~\cite{Barabash:2011mf}. Although $0\nu\beta\beta$ experiments were still performed in the 1960s and 1970s, the main goal was to test the lepton number conservation law. Only until the early 1980s the search for $0\nu\beta\beta$ embraced its ``Renaissance" period~\cite{Barabash:2011mf}, mainly because the grand unified theories developed around that time can naturally generate a Majorana mass term for neutrinos, and neutrinos with tens of eV masses were thought to be possible for being a good dark matter candidate in cosmology. 

In the current ``contemporary ages"~\cite{Barabash:2011mf}, the observation of neutrino oscillation, which indicates that neutrinos have masses, really boomed this field. Several experiments have been built, or are under construction, and dozens of proposals are made for future investigations. Such dedicated experimental efforts may enable us to extract rich physics results from $0\nu\beta\beta$ in a near future. For example,

\begin{itemize}

\item Pinning down the nature of neutrinos. According to Schechter-Valle theorem~\cite{Schechter:1981bd}, the observation of $0\nu\beta\beta$ would imply a Majorana mass term for neutrinos, proving the Majorana nature of neutrinos.  

\item Constraining the absolute neutrino mass scale and lepton CP-violating phases. This is because that if $0\nu\beta\beta$ is mediated by the light neutrino Majorana mass term, the decay half-life $T_{1/2}^{0\nu}$ can then be written as
\begin{eqnarray}\label{eq:rate}
\left ( T_{1/2}^{0\nu} \right)^{-1} = G_{0\nu}^{} ~|\mathcal{M}_{0\nu}^{}|^2 ~m_{\beta\beta}^2 \; ,
\end{eqnarray}
where $G^{}_{0\nu}$ and $\mathcal{M}_{0\nu}^{}$ stand for the phase-space factor \cite{Kotila:2012zza} and the nuclear matrix element (NME), respectively, and $m_{\beta\beta}^{}$ is the so-called effective neutrino mass. Although there still exist large uncertainties in the calculation of NME~\cite{Engel:2015wha}, in principle, one can obtain the former two accurately within nuclear theory. Then, the measured limit on the half-life $T_{1/2}^{0\nu}$ would directly yield a constraint on $m_{\beta\beta}^{}$, whose definition is given by
\begin{eqnarray} \label{eq:mbb}
m_{\beta\beta}^{} = |m_1^{}\cos^2\theta_{12}\cos^2\theta_{13} e^{2\mathrm{i} \varphi_1^{}} + m_2^{}\sin^2\theta_{12}\cos^2\theta_{13}  e^{2\mathrm{i} \varphi_2^{}} + m_3^{} \sin^2\theta_{13}| \; , \nonumber \\
\end{eqnarray}
where $m_i^{}$'s are neutrino masses, and $\theta_{ij}^{}$'s and $\varphi_{i}^{}$'s are the mixing angles and CP-violating phases in the lepton mixing matrix, respectively. 

\item Determining the neutrino mass ordering. This is due to the presence of a lower bound for the above effective neutrino mass $m_{\beta\beta}^{}$ in the inverted neutrino mass ordering, according to the current global fit results~\cite{Gonzalez-Garcia:2014bfa}. 

\item Restricting new physics contributions. As $0\nu\beta\beta$ is a rare process, any new physics that contributes to it should be constrained by the obtained data. 

\end{itemize}

In this talk we focus on the question of determining neutrino mass ordering in the next generation $0\nu\beta\beta$ experiments~\cite{Zhang:2015kaa}. Our goal is to provide a quantitative description of experimental power to distinguish between normal neutrino mass ordering (NO) and inverted neutrino mass ordering (IO). For simplicity, we only consider $^{76}$Ge-type experiments, as their background is found to be nearly flat in the signal region. Moreover, we adopt Bayesian analysis as our statistical method. 

\section{Description of $0\nu\beta\beta$ Experiments} \label{sec:exps}

Since we are discussing the future generation $0\nu\beta\beta$ experiments, it would be beneficial to review what the indispensable ingredients are, if one wants to build an ultimate $0\nu\beta\beta$ experiment~\cite{GomezCadenas:2011it}. 

\begin{table}
\tbl{Simple classification of different generations of $0\nu\beta\beta$ experiments.}{
\begin{tabular}{c | c | c | c}
\hline
\hline
Generation & Size & BI(counts/keV/kg/yr) & Half-life sensitivity \\
\hline
Previous & $\sim 10~{\rm kg}$ & $\sim 10^{-2}_{}$ & $\sim 10^{25}_{} ~{\rm yr}$ \\
Current & $\sim 100~{\rm kg}$ & $\sim 10^{-3}_{}$ & $\sim 10^{26 \sim 27}_{} ~{\rm yr}$ \\
Next & $\sim {\rm ton}$ & $\sim 10^{-4}_{}$ & $\sim 10^{28}_{} ~{\rm yr}$ \\
\hline
\hline
\end{tabular}}
\label{tb:classification}
\end{table}

\begin{itemize}

\item Large amount of $\beta\beta$-decaying isotopes. The number of signal events $N^{0\nu}_{}$ is directly related to the fiducial mass of $\beta\beta$-decaying isotopes. Taking $^{76}$Ge for an example, we have
\begin{eqnarray} \label{eq:Ge}
N^{0\nu}_{} = \ln 2 \cdot N_{\rm A}^{} \cdot\frac{ \mathcal{E} \cdot \epsilon}{M_{\mathrm{Ge}}^{} \cdot T_{1/2}^{0\nu}} \; ,
\end{eqnarray}
where $N_{\rm A}^{}$ is the Avogadro's constant, $M_{\mathrm{Ge}} = 75.6~{\rm g/mol}$ is the molar mass of $^{76}$Ge, and $\mathcal{E}$ and $\epsilon$ are the exposure and detection efficiency, respectively. Since the exposure $\mathcal{E}$ is defined as the product of the fiducial mass and the experiment operating time, increasing the mass of isotopes is helpful to reduce the detection time. In Table \ref{tb:classification} we list the typical sizes of the fiducial mass in the previous, current and next generation $0\nu\beta\beta$ experiments. 

\item Good energy resolution. Ideally, all the above signal events should show up at the Q-value of the corresponding $0\nu\beta\beta$ process. For $^{76}$Ge, such a Q-value is 2039 keV. However, because of a finite energy resolution, these signal events spread out and can be easily contaminated by background events, especially those from the intrinsic $2\nu\beta\beta$ process. Because $^{76}$Ge-type experiments employ the ``detector = studied substance" scheme, a very high energy resolution can be reached. In this work we take FWHM (full width half maximum) as 3 keV.

\item Low background. It was found that reducing the level of background is very helpful to improve the half-life sensitivity~\cite{GomezCadenas:2011it}. Moreover, if possible, one should pursue the zero background limit, so that a single $0\nu\beta\beta$ event would be enough to pin down the Majorana nature of neutrinos. For the next generation $^{76}$Ge-type experiments we will choose ${\rm BI} = 10^{-4}_{}$~counts/keV/kg/yr, with BI standing for background index. BI's of other generations of $0\nu\beta\beta$ experiments are also given in Table \ref{tb:classification}.

\item Good detection efficiency. $0\nu\beta\beta$ happens so rare that any signal event would be precious. Here we take the typical value of $\epsilon = 0.65$ for $^{76}$Ge-type experiments. 

\item Good scalability. By scalability we mean the ability to scale the experimental set-up to a larger size. This is very important for building a large $0\nu\beta\beta$ experiment stage by stage. 

\end{itemize}

%Unfortunately, none of current $0\nu\beta\beta$ experiments scores highest on all the above aspects. 

Currently, we have two commissioning $^{76}$Ge-type $0\nu\beta\beta$ experiments, i.e., G{\scriptsize ERDA} and M{\scriptsize AJORANA} D{\scriptsize EMONSTRATOR} \cite{Abgrall:2013rze}. In the first phase of G{\scriptsize ERDA}, a lower limit of $T_{1/2}^{0\nu} > 2.1\times 10^{25}$ yr at $90\%$ confidence level was reported for an exposure of 21.6 kg~$\cdot$~yr and a BI about $10^{-2}$~\cite{Agostini:2013mzu}. The G{\scriptsize ERDA} Phase-II and M{\scriptsize AJORANA} D{\scriptsize EMONSTRATOR} are expected to increase the exposure to about 200 kg~$\cdot$~yr each, and to reduce BI to the $10^{-3}$ level at the same time.  More excitingly, these two collaborations are also discussing the possibility of building a future large scale $^{76}$Ge (LSGe) experiment together~\cite{LSGe}, which may eventually reach an exposure of around $10^4$ kg~$\cdot$~yr, and ${\rm BI} = 10^{-4}$. It is such a possibility that motivates us to perform a detailed statistical analysis, so as to find out at which level the currently undetermined neutrino mass ordering can be resolved then.

\section{Statistical Determination of Neutrino Mass Ordering}

\subsection{Generating Pseudo-data Sample}

\begin{figure}
\begin{center}
\includegraphics[scale=0.75]{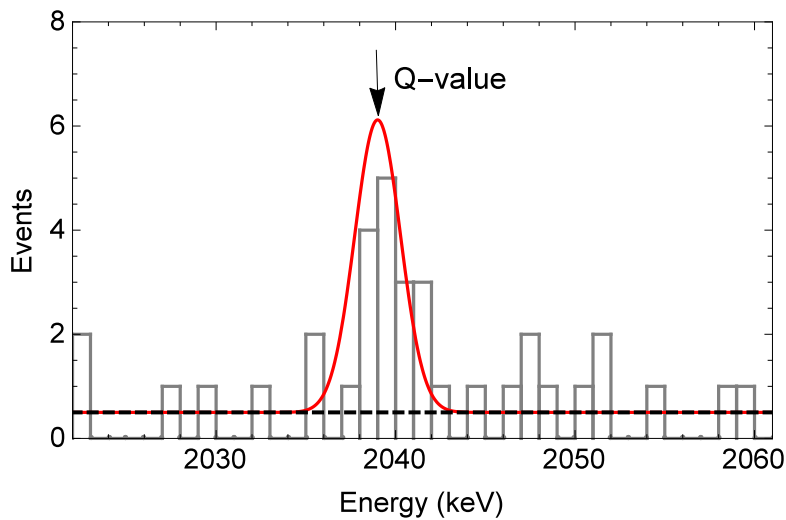}
\end{center}
\vspace{-0.5cm}
\caption{A simulated event spectrum of $0 \nu \beta \beta$ for the $^{76}{\rm Ge}$-based experiment, where the red solid curve represents the expected Gaussian distribution of signal events on top of a constant background and typical values of the half life $T^{0\nu}_{1/2} = 10^{25}~{\rm yr}$, the exposure ${\cal E} = 50~{\rm kg} \cdot {\rm yr}$, the energy resolution ${\rm FWHM} = 3~{\rm keV}$, and the efficiency $\epsilon = 0.65$ have been used. The black-dashed horizontal line corresponds to the background with ${\rm BI} = 10^{-2}$. The gray histograms stand for the total number of both signal and background events, which is randomly generated according to the Poissonian distribution in each energy bin. Taken from Ref.~[6]~\cite{Zhang:2015kaa}.}
\label{fg:spectrum}
\end{figure}

We begin with generating pseudo-data samples that are to be used in the later statistical analysis. Following G{\scriptsize ERDA}'s analysis on its first phase data \cite{Agostini:2013mzu}, we consider a region of spectrum that spans from 2022 keV to 2061 keV with a bin size of 1 keV. See Fig.~\ref{fg:spectrum} for an example of the simulated event spectrum. Red curve represents the total un-binned signal and background events, while the binned results are given by the gray histogram, where the Poisson statistics is assumed. 

In Ref.~[6]~\cite{Zhang:2015kaa}, we indeed adopt the above generated spectrum, which incorporates statistical fluctuation, to perform analysis on the exclusion limit and discovery potential in the upcoming $^{76}$Ge-type experiments. However, for the current discussion on the determination of neutrino mass orderings, using the above statistical-fluctuation-included spectrum would be very time-consuming. This is because for each set of input parameters, one needs to generate and then analyze hundreds of pseudo-data samples so as to obtain enough statistics. 

For simplicity, here we instead take the so-called Asimov data set~\cite{Cowan:2010js}, which is obtained by simply assuming the \emph{expected} number of events for each bin in the spectrum. In terms of the example given in Fig.~\ref{fg:spectrum}, the Asimov data set just means the events under the red curve. Apparently, no statistical fluctuations are included in this Asimov data set. However, it was found that the analysis of the Asimov data yields a good approximation to the \emph{median} projection of experiments~\cite{Cowan:2010js}. For our current discussion such a median projection is sufficient.

\subsection{Bayesian Analysis}

We adopt Bayesian analysis as our statistical method. Bayesian analysis resides on the well-known Bayes' theorem, and describes the degree of belief in a certain hypothesis $\mathcal{H}^{}_i$, given the data set $\mathcal{D}$. Specifically, Bayes' theorem states that
\begin{eqnarray} \label{eq:bayes}
\mathrm{Pr}(\mathcal{H}^{}_i|\mathcal{D}) = \frac{\mathrm{Pr}(\mathcal{D}|\mathcal{H}^{}_i) ~\mathrm{Pr}(\mathcal{H}^{}_i)}{\mathrm{Pr}(\mathcal{D})} \; ,
\end{eqnarray}
where $\mathrm{Pr}(\mathcal{H}^{}_i)$ and $\mathrm{Pr}(\mathcal{H}^{}_i|\mathcal{D})$ are the prior and posterior probabilities of the hypothesis $\mathcal{H}^{}_i$, respectively. The probability of obtaining the data $\mathcal{D}$, given the hypothesis $\mathcal{H}_i$ to be true, is denoted as $\mathrm{Pr}(\mathcal{D}|\mathcal{H}_i)$, and is also called the evidence $\mathcal{Z}_i$. Lastly, $\mathrm{Pr}(\mathcal{D})$ stands for the overall probability of observing the data $\mathcal{D}$, and it is equal to $\sum_{i}^{} \mathrm{Pr}(\mathcal{D}|\mathcal{H}_i) \mathrm{Pr}(\mathcal{H}_i)$, because of the normalization condition $\sum_{i}^{} \mathrm{Pr}(\mathcal{H}_i|{\cal D}) = 1$.

Following the above formalism, we take our two competing hypotheses as $\mathcal{H}_{\rm NO}^{}$ and $\mathcal{H}_{\rm IO}^{}$, and the above discussed pseudo-data samples would be our data set $\mathcal{D}$. To find out the more favorable hypothesis, we compute the following posterior odds 
\begin{eqnarray}
\frac{\mathrm{Pr}(\mathcal{H}_{\rm NO}^{}|\mathcal{D})}{\mathrm{Pr}(\mathcal{H}_{\rm IO}^{}|\mathcal{D})} = \frac{\mathcal{Z}_{\rm NO}^{}}{\mathcal{Z}_{\rm IO}^{}} \frac{\mathrm{Pr}(\mathcal{H}_{\rm NO}^{})}{\mathrm{Pr}(\mathcal{H}_{\rm IO}^{})} \; ,
\end{eqnarray}
where the ratio of evidences ${\cal B} \equiv \mathcal{Z}_{\rm NO}^{}/\mathcal{Z}_{\rm IO}^{}$ is termed Bayes factor. If \emph{a prior} we have no preference for a particular mass ordering, the above posterior odds is then directly reflected by the Bayes factor $\mathcal{B}$. Furthermore, to interpret the value of this posterior odds or the Bayes factor, we adopt the Jeffreys scale~\cite{Jeffreys} given in Table~\ref{tb:Jeffreys}. 

\begin{table}
\tbl{The Jeffreys scale used for the statistical interpretation of Bayes factors and posterior odds~\cite{Jeffreys_mod, Trotta:2008qt}.}{
\begin{tabular}{l | l | l | l}
\hline
\hline
$\left|\ln(\text{odds})\right|$ & Odds & Probability & Interpretation \\
\hline
$< 1.0$ & $\lesssim 3 : 1$ & $\lesssim 75.0\%$ & Inconclusive \\
$1.0$ & $\simeq 3 : 1$ & $\simeq 75.0\%$ & Weak evidence\\
$2.5$ & $\simeq 12 : 1$ & $\simeq 92.3\%$ & Moderate evidence \\
$5.0$ & $\simeq 150 : 1$ & $\simeq 99.3\%$ & Strong evidence\\
\hline
\hline
\end{tabular}}
\label{tb:Jeffreys}
\end{table}

The computation of $\mathcal{Z}_{\rm NO}^{}$ is performed via $ \mathcal{Z}_{\rm NO}= \int {\rm Pr}({\cal D}|\Theta, {\rm NO}) \pi (\Theta) \mathrm{d}^{\rm N}_{}\Theta $ (similar computation for $\mathcal{Z}_{\rm IO}^{}$). $\Theta$ stands for the parameters in the hypothesis $\mathcal{H}_{\rm NO}^{}$, and here they are three lepton mixing angles $\theta^{}_{ij}$, two CP-violating phases $\{\varphi^{}_1, \varphi^{}_2\}$, two neutrino mass-squared differences and the lightest neutrino mass $m_0^{}$. Their prior probability distributions are collectively denoted as $\pi(\Theta)$. For the mixing angles and mass-squared differences, we assume Gaussian priors with the central values and $1\sigma$ errors taken from Ref.[5]~\cite{Gonzalez-Garcia:2014bfa}. For CP-violating phases, uniform priors are chosen, and for the lightest neutrino mass $m_0^{}$ we adopt two different priors, i.e., a uniform prior within $[0, 0.2]~\text{eV}$ and a logarithmic prior on $[10^{-5}, 0.2]~\text{eV}$. Finally, we obtain ${\rm Pr}({\cal D}|\Theta, {\rm NO})$ by fitting the pseudo-data set with the above sampled parameters in the realm of $\mathcal{H}_{\rm NO}^{}$.

\subsection{Numerical Results}

\begin{figure}
\begin{center}
\includegraphics[scale=0.7]{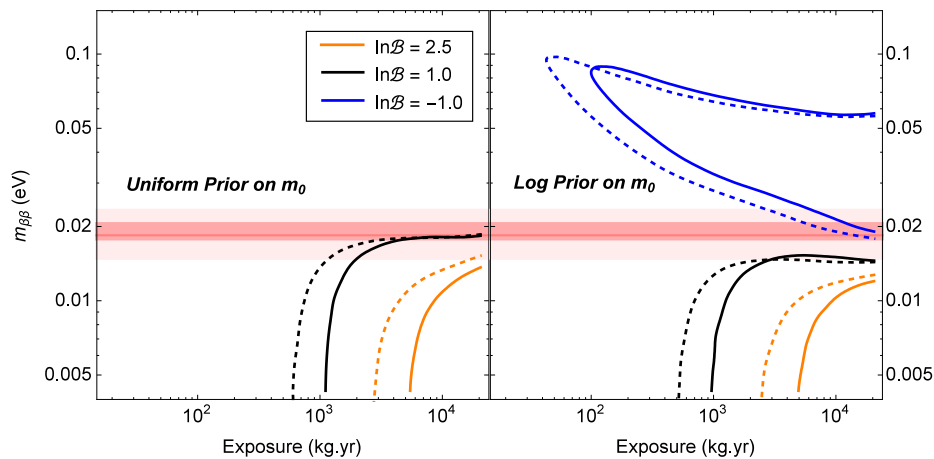}
\end{center}
\vspace{-0.5cm}
\caption{Contours of the Bayes factor $\ln{\cal B} \equiv \ln({\cal Z}^{}_{\rm NO}/{\cal Z}^{}_{\rm IO})$ calculated by comparing between NO with IO, where the thick solid (dashed) curves correspond to the low NME ${\cal M}^{}_{0\nu} = 4.6$ (the high NME ${\cal M}^{}_{0\nu} = 5.8$). The orange, black and blue curves represent $\ln({\cal B}) = 2.5$, $1$ and $-1$, respectively. Taken from Ref.~[6]~\cite{Zhang:2015kaa}.}
\label{fg:NH_IH}
\end{figure}

We now present the finally obtained numerical results on the discrimination of two neutrino mass orderings in the next generation $^{76}$Ge $0\nu\beta\beta$ experiments, see Fig.~\ref{fg:NH_IH}.

In Fig.~\ref{fg:NH_IH}, the left and right panels correspond to the uniform and logarithmic priors on the lightest neutrino mass $m_0^{}$, respectively. Given a true value of $m^{}_{\beta \beta}$ and an exposure ${\cal E}$, we can calculate the Bayes factor ${\cal B}$, and show its contours as orange, black and blue curves for $\ln({\cal B}) = 2.5$, $1$ and $-1$, respectively. Thick solid and dashed curves distinguish two different choices of NME, i.e., ${\cal M}^{}_{0\nu} = 4.6$ and $5.8$ for the former and later, respectively. For reference, we also plot the lower bound of ${m^{}_{\beta\beta}}$ in IO, i.e., the red horizontal line represents the result calculated by using the best-fit values of neutrino mixing parameters, while the dark (light) band is for the result by using $1\sigma$ ($3\sigma$) ranges.

From Fig.~\ref{fg:NH_IH}, we then make the following observations:
\begin{itemize}

\item An exposure of at least $500$ (or $2500$) kg~$\cdot$~yr is needed in order to report a weak (or moderate) evidence for NO. According to the Jeffreys scale in Table~\ref{tb:Jeffreys}, the weak and strong evidence should be understood as a degree of belief of $75.0\%$ and $92.3\%$, respectively.

\item In the next generation of $^{76}$Ge-based $0\nu\beta\beta$ experiments with an exposure of $10^4~{\rm kg}\cdot{\rm yr}$, we are only able to reach a value of $\ln(\mathcal{B}) \gtrsim 2.5$, indicating a moderate evidence for NO. To obtain a strong evidence ($\ln(\mathcal{B}) = 5)$, one needs to further increase the exposure or reduce the level of background. 

\end{itemize}

\section{Summary and Conclusion}

The question of whether neutrino is its own anti-particle, namely, the nature of neutrino, leads to a long history of searching for $0\nu\beta\beta$ experimentally. In the course of such a dedicated search, one also realizes that studying $0\nu\beta\beta$ may help us distinguish the currently unresolved neutrino mass ordering. Motivated by the latter fact, we perform a detailed quantitative analysis, aiming at finding out at which statistical level one can exclude the inverted neutrino mass ordering in the next generation $^{76}$Ge-type $0\nu\beta\beta$ experiments. Our finding indicates that the next generation $^{76}$Ge-type $0\nu\beta\beta$ experiments indeed has some sensitivity to the discrimination of neutrino mass orderings, however, not in a very decisive manner.

\section*{Acknowledgement}

The author thanks Prof. Shun Zhou for the fruitful collaboration on the subject. He is also grateful to Prof. Harald Fritzsch for the kind invitation, and local organizers for the warm hospitality. This work was supported in part by the Innovation Program of the Institute of High Energy Physics under Grant No. Y4515570U1.

\end{document}